\documentclass{article}

\IfFileExists{microtype.sty}{\usepackage{microtype}}{}
\IfFileExists{graphicx.sty}{\usepackage{graphicx}}{}
\IfFileExists{subcaption.sty}{\usepackage{subcaption}}{}
\IfFileExists{booktabs.sty}{\usepackage{booktabs}}{}
\IfFileExists{float.sty}{\usepackage{float}}{}
\IfFileExists{xurl.sty}{\usepackage{xurl}}{}
\IfFileExists{hyperref.sty}{\usepackage{hyperref}}{}

\usepackage[preprint]{icml2026}

\makeatletter
\renewcommand{\ICML@preprint}{%
  \textit{Accepted to the AI4Law Workshop at ICML 2026.}%
}
\makeatother

\IfFileExists{amsmath.sty}{\usepackage{amsmath}}{}
\IfFileExists{amssymb.sty}{\usepackage{amssymb}}{}
\IfFileExists{mathtools.sty}{\usepackage{mathtools}}{}
\IfFileExists{amsthm.sty}{\usepackage{amsthm}}{}
\IfFileExists{cleveref.sty}{\usepackage[capitalize,noabbrev]{cleveref}}{}

\graphicspath{{./}{figures/}}

\newcommand{\inlinepaperfigure}[4]{%
  \begin{figure}[H]
    \centering
    \IfFileExists{#1}{%
      \includegraphics[width=#2\linewidth]{#1}%
    }{%
      \fbox{\parbox{0.92\linewidth}{\centering Missing figure file: \texttt{\detokenize{#1}}}}%
    }%
    \caption{#3}
    \label{#4}
  \end{figure}
}

\icmltitlerunning{Legal Reasoning Is Not Lawyering}

\begin{document}

\twocolumn[
  \icmltitle{Legal Reasoning Is Not Lawyering: Rethinking Legal Benchmarks for Pro Se Access to Justice}

  \icmlsetsymbol{equal}{*}

  \begin{icmlauthorlist}
    \icmlauthor{Andrew Lou}{equal,yale}
    \icmlauthor{David Shin}{equal,yale}
  \end{icmlauthorlist}

  \icmlaffiliation{yale}{Yale Law School, USA}

  \icmlcorrespondingauthor{Andrew Lou}{andrew.lou@yale.edu}
  \icmlcorrespondingauthor{David Shin}{d.shin@yale.edu}

  \icmlkeywords{legal reasoning, legal AI, access to justice, benchmarks, pro se litigation}

  \vskip 0.3in
]

\printAffiliationsAndNotice{\icmlEqualContribution }

\begin{abstract}
Legal AI benchmark research will frequently invoke the assumption that large language models can improve access to justice, including for people who cannot access lawyers in order to understand and exercise their legal rights \citep{guha2023legalbench}. However, we argue that current benchmarks are not equipped to support this assumption because they evaluate legal reasoning over inputs that have already been preprocessed by legal experts, which measures the \textit{upper bound} of model performance. Access to justice depends on a \textit{lower bound}: how models perform when inputs come from pro se litigants, whose prompts may contain noisy narratives, buried facts, omissions, folk-legal assumptions, and surface-level errors. These degradations are comparable to the conditions under which LLMs have been shown to degrade in the general ML literature, such as long context sensitivity, underspecification, hallucination, and typographical perturbations. We connect evidence from pro se literature with this body of ML research and present a small perturbation experiment on LEXam, a leading legal benchmark, to illustrate the gap between the two bounds. If models continue to focus on improvements based on current benchmarks that only measure the upper bound, this gap will continue to remain hidden or even widen. We conclude by calling for legal benchmarks that directly measure robustness under pro se-like inputs so that access to justice claims about legal AI can become empirically testable.

\end{abstract}

\section{Introduction}

Legal benchmarks have made substantial progress in measuring how well general-purpose models reason about legal questions. To name a few, LegalBench evaluates issue spotting and rule-application across a wide taxonomy of tasks, LEXam tests long-form legal reasoning derived from real law school examinations, and LegalBench-RAG measures end-to-end RAG retrieval performance \citep{guha2023legalbench,fan2025lexam,pipitone2024legalbenchrag}. Using this rapidly expanding infrastructure, researchers are noticing continuing improvements with each new frontier model, prompting questions as to how close AI is to high-stakes deployment in the real world. A common presumption is that more powerful models will naturally lead to improved access to justice by giving unrepresented litigants a tool to better understand and exercise their legal rights \citep{toycronin2022tighten,lsc2022justicegap}.

However, we argue that the current legal benchmark infrastructure is ill-equipped to assess the readiness of AI to improve access to justice\footnote{Access to justice is not a universally agreed-upon set of policy objectives. Nonetheless, we will continually refer to access to justice in this paper in reference to unrepresented claimants that attempt to access their legal remedies through the legal system of their jurisdiction.} because it does not evaluate the correct capabilities required to draw conclusions for those applications. Thus, if researchers continue to compete for rankings on current legal benchmarks, it is not obvious access to justice will naturally improve as a byproduct. This is not a matter of incomplete coverage, in which case adding more domains becomes the solution. Rather, current legal benchmarks are measuring the distinct skill of legal reasoning, which we differentiate from the more general skill of lawyering.

Legal reasoning, as measured by current benchmarks, is the application of legal doctrine to already well-formed legal questions. Every legal benchmark we are aware of evaluates models on lawyer mediated inputs \citep{guha2023legalbench,fan2025lexam,pipitone2024legalbenchrag}. The benchmark administrator presents the model with prompts that have facts in a legally legible order, the irrelevant details already pruned, and the procedural posture clarified. Thus, by the time a benchmark prompt reaches a model, a substantial amount of cognitive labor by legal experts has already been performed.

However, this preprocessing step should not be taken for granted. The ability to extract relevant legal facts from a noisy narrative and present them cogently along with the appropriate procedural posture is precisely the hallmark of a great lawyer. For purposes of this paper, we call this lawyering. Most unrepresented legal claimants, commonly known as pro se in the legal profession, cannot be expected to translate an emotionally loaded chronology of facts (or lack thereof) into an industry standard pleading fit to proceed \citep{levy2018empirical,stienstra2011assistance}.

Therefore, while current benchmarks reflect a sensible design choice for measuring legal reasoning capability in the abstract, they are poor proxies for measuring how a model will improve equitable access to justice for legally vulnerable populations. Legal benchmarks only test for legal reasoning under optimal conditions, which forms the upper bound on frontier model capabilities. But it is in fact the lower bound of model performance that provides a more realistic estimate of how frontier models can systematically improve legal access for unrepresented litigants.

The gap between current benchmark design and access to justice conditions is not speculative. It rests on two independently established premises. First, unsurprisingly, pro se submissions differ from lawyer-mediated submissions along documentable dimensions. Second, general large language model performance is known to degrade along analogous perturbations.

In combining these two presumptions, we find that while the relevant machine learning literature has established what happens when input quality degrades in general use cases, the legal benchmark literature has not adopted similar robustness analysis as standard practice \citep{guha2023legalbench,fan2025lexam,pipitone2024legalbenchrag}, which has increasingly larger ramifications for users the closer they are to the lower-bound scenario. Counterintuitively, this means that as frontier models improve solely based on current benchmarks, the inequality in legal resources between unrepresented litigants and represented litigants may compound instead of subside, since represented litigants can get more out of legal AI models whereas pro se litigants can only extract lower quality output. This is because the upper bound of performance can rise without the lower bound rising commensurately or even at the expense of the lower bound. And since benchmarks do not measure the lower bound at all, developers cannot detect this, leaving the user uncertain as to how large or small that gap has changed. Additionally, it is possible that model robustness to input quality can vary significantly across different frontier models, and the best model for raw legal reasoning may not be the best model for a pro se litigant.

We suspect this is symptomatic of a broader pattern of how machine learning research approaches benchmarks, where researchers focus more on clean inputs as opposed deployment case inputs. However, this pattern is unusually consequential for the legal industry because the divergence between benchmark and deployment can be closely tied to an access to justice spectrum. Moreover, continued focus on the upper bound of frontier model capability bakes in assumptions of a lawyer-centric legal system into future legal AI development. Ironically, the population that bears the cost of the divergence is also the population that legal AI researchers are often most interested in when invoking a justification for their work.

This paper ultimately aims to contribute to the growing legal benchmark literature by arguing that current leading benchmarks cannot be interpreted to draw conclusions about access to justice because they do not evaluate the appropriate variable. We develop this argument through the following parts.  Section 2 surveys the unique challenges pro se litigants face and documents how their interactions with the legal system differ from those of represented parties. Section 3 examines the literature on the established limitations of LLMs when inputs deviate from benchmark-typical conditions. Section 4 brings the two previous sections into contact by observing what current benchmarks are measuring, what they are missing in connection to the foreseeable limitations of pro se litigation, and how the gap between measured and access to justice-relevant performance can widen as models improve. In Section 5, we make these arguments concrete with an experimental primer that uses a sample of multiple-choice questions from LEXam, transforming a sample of multiple-choice questions along two forms of degradations in line with the pro se literature to show deteriorating model performance across three models. We conclude in Section 6 with suggestions for how legal AI evaluation should evolve.

\section{Pro Se Litigation}

\subsection{Scale of Self-Representation}

The scale of self-representation in the American legal system is substantial. In federal civil litigation, 27\% of all civil cases from 2000 to 2019 included at least one pro se plaintiff or defendant \citep{uscourts2021prose}. In state courts, which handle the vast majority of civil cases, approximately three-quarters of the twenty million cases filed annually involve at least one self-represented party \citep{hannafordagor2015landscape,engstrom2024making}. In specific dockets, the rates approach universality. 98\% of tenants in New York eviction proceedings appeared without counsel in 2013 \citep{taskforce2010expand}.

Although the reasons people proceed as pro se may be heterogeneous, pro se litigants are routinely concentrated among lower-income parties facing all kinds of legal problems ranging from eviction to wage theft \citep{lsc2022justicegap}. What this population faces upon entering the legal system is a tedious array of procedures and requirements that were not designed with them in mind. Civil procedure assumes a party who can draft pleadings that comply with form requirements while substantive law assumes a party that can identify relevant doctrine. None of these capabilities are part of lay legal literacy.

\subsection{Noted Deficiencies}

The consequences of such deficiencies are easy to identify. A Federal Judicial Center survey of district court chief judges and clerks of court identified some of the recurring difficulties found in pro se submissions: physically or formally hard to read, omitted yet material information, and salient facts buried in an unreadable narrative, often in unhelpful chronological order \citep{stienstra2011assistance}.

Similarly, a controlled experimental study using a mock online court portal compared claims drafted by lay participants to claims drafted by trained lawyers from the same underlying fact patterns and found that lay drafters were less effective at reporting legally relevant details, that they sometimes reported such details without clear purpose or in ways that confused the reader. Retired judges evaluating blinded submissions rated the overall quality of lay-drafted claims as lower than lawyer-drafted claims, even as the study identified outliers in both directions \citep{toycronin2022tighten}.

Research on small-claims litigation, where pro se representation is the norm, shows similar findings. O'Barr and Conley's ethnographic work on small-claims narratives found that the less formal setting of small-claims courts allows lay litigants to feel more comfortable producing emotionally compelling narratives, but those stories nonetheless diverge from the legally relevant arguments that adjudication requires \citep{obarr1985litigant,conley1990rules}.

\subsection{Documented AI Usage}

Lack of such procedural knowledge presents a barrier for pro se litigants that is distinct to legal reasoning. Another problem arises when they try to fill in that obstacle with general AI, whose barrier to use is continually lowering.

This concern is not hypothetical. Courts have begun to encounter pro se filings that bear the marks of frontier model assistance including fabricated case citations and confidently stated (yet incorrect) propositions that are not evaluated critically. Recent rulings have addressed both how to handle such filings and the broader question of how AI-assisted pro se litigation is reshaping court dockets \citep{fisherphillips2026playbook,polson2026chatgpt}. Such commentary is an indicator that benchmark performance is conflating lawyering and legal reasoning skills in the perceptions of lay litigants and leading to more adverse consequences.

\section{LLM Sensitivities}

The first premise of our argument, that pro se submissions differ significantly from lawyer-mediated submissions, is established in the prior section. The second premise is that these specific differences are very comparable to the types of conditions under which large language models have been shown to perform poorly. Rather than survey the substantial body of computer science literature already examining the sensitivity of LLMs to decreased input quality, prompt design and contextual framing, this paper will focus on four particular degradations that can be mapped onto the specific deficiencies catalogued in the pro se literature.

\subsection{Long Context Processing}

Solving real-world legal questions often requires intensive fact collection. For large language models to take on this task, models need long context windows that are capable of processing hundreds, if not thousands, of tokens. While many new models are able to handle long context prompts, long prompts still put significant strain on its reasoning capabilities. Even when a model can accommodate such submissions, what models do with information inside those windows is a separate question. By varying placement of relevant information within a long prompt, Liu et al. found that performance is highest when relevant information appears at the beginning or end of the context. However, performance degraded substantially when the same information was placed in the middle and the effect persisted even on models explicitly trained for long contexts \citep{liu2024lost}. As the previous section noted, lay drafters tend to bury salient facts in narrative ordering which the FJC survey of chief judges identified as one of the most common pro se deficiencies \citep{stienstra2011assistance}. The combination of these findings suggests that even when a pro se prompt contains everything a model would need, the model's ability to use that information depends on where in the narrative it appears. Lawyering, in this respect, is not only a matter of getting the right facts but emphasizing them and noticing when they are not properly accounted for in the output.

\subsection{Omission and Abstention}

What if relevant information is omitted? If a pro se client doesn't understand which facts matter, they may fail to include it in the prompt altogether. Underspecification occurs when a user omits information necessary for the requested output. In such cases, models ideally would abstain from answering with a definitive answer or ask a clarifying question. Unfortunately, research has found reasoning LLMs consistently faulty at abstention. Yang et al. measured this behavior on a curated set of underspecified prompts and found that, LLMs will silently supply assumed values for omitted facts and proceed to an answer 41.1\% of the time \citep{yang2025prompts}. To make matters worse, underspecified prompts were twice as likely to regress in performance across model or prompt changes compared to fully specified prompts, and accuracy would drop up to twenty percentage points relative to the fully specified counterpart \citep{yang2025prompts}. Further research by Kirichenko et al. went as far as to claim that abstention is "an unsolved problem" across the frontier and that increasing model scale provides little marginal benefit. Alarmingly, they found that reasoning fine-tuning, which produces measurable accuracy gains on reasoning benchmarks, degrades abstention by an average of 24\%, even in domains on which reasoning models are explicitly trained \citep{kirichenko2025abstentionbench}.

\subsection{Hallucinations}

When LLMs fail to abstain when facts are sparse and underspecified, hallucination is likely to occur. When a study from 2024 evaluated public-facing LLMs on questions about federal court cases, they reported hallucination rates between 58\% (Chat GPT-4) and 88\% (Llama 2) \citep{dahl2024large}. While LLMs have certainly improved by leaps and bounds since the 2024 report, hallucinations in court filings by prominent law firms as recent as April 2026 indicate that the hallucination problem is far from solved, even when sophisticated parties are using AI \citep{polson2026chatgpt}. And when pro se users are unable to use legal terminology in describing their claims and instead resort to popular legal misconceptions or lay terminology \citep{stienstra2011assistance,toycronin2022tighten}, it is plausible to assume that the ambiguity generated may give rise to even more frequent legal hallucinations. The most concerning outcome, which is likely also the most probable, is when the user's mistaken legal premises are accepted and used as the foundation of a model's output.

\subsection{Character Typos}
The perturbations catalogued thus far involve substantive variations. A separate body of research has measured how LLMs respond to surface-level changes in input quality. Studies such as PromptRobust reported that even minor perturbations such as typos or grammatical errors can produce performance drops of roughly 20\% to 33\% across an array of tasks \citep{zhu2023promptbenchrobust}. Additional studies combining typographical and lexical perturbations have shown non-additive performance drops \citep{zhu2023promptbenchrobust}, while other findings have even shown that extra polite requests, by virtue of consuming more context tokens, also decrease the quality of model output \citep{dobariya2025mindtone}.

This pattern across the ML literature is sufficiently consistent with the prediction that frontier legal AI performance under pro se input conditions will diverge substantially from performance under lawyer-mediated conditions. This does not require us to discover a new phenomenon. It requires only that we measure, in the legal domain, what general ML research has already established holds more broadly. We turn next to the implications of bringing the previous two sections into contact and what that interaction suggests about the limits of inferring access to justice readiness from current benchmark scores.

\section{Implications}

\subsection{What Current Benchmarks are Measuring}

When a frontier model scores well on LegalBench, LEXam, or LegalBench-RAG, what has been measured is the model's capacity to perform legal reasoning on inputs pre-processed with lawyering \citep{guha2023legalbench,fan2025lexam,pipitone2024legalbenchrag}. Thus, a model's performance on such benchmarks is only helpful in answering whether it can reason about clear legal questions. This is a meaningful question and current benchmarks answer it well. However, it is not the question that determines whether AI improves access to justice.

For deployment settings where upstream lawyering is performed, such as when a lawyer is using a model for her client, performance on current benchmarks may serve as a reasonable proxy for deployment practice. However, for settings in which there is no lawyer upstream, such as the pro se setting, current benchmarks measure a quantity whose relationship to deployment performance is not established. The size of this gap may vary depending on which dimensions in section 2 and 3 we are referring to.

\subsection{The Widening Gap}

But as long as a gap exists, AI tools may inadvertently increase inequity rather than decreasing it in the legal field. And if legal benchmarks incentive models to maximize their legal reasoning capacity over time, we may see a world in which the upper bound of legal reasoning rises while the lower bound of legal reasoning stays stagnant. Counterintuitively, then, legal AI may help grow instead of narrow the legal resources disparity between unrepresented litigants and represented litigants.

Theoretically, this makes sense. The gains from improved legal reasoning will flow most directly to users who can supply lawyer-mediated inputs. A pro se litigant cannot supply these things, by virtue of not having the legal training that would let them know what to supply. Improvements in reasoning capability do not address this bottleneck.

There is also a more specific mechanism in the training regime that produces this widening effect. Kirichenko et al. find that reasoning fine-tuning, which produces the gains visible on legal benchmarks, degrades abstention by an average of 24\%, even in domains the reasoning training targets \citep{kirichenko2025abstentionbench}. Joren et al. find that proprietary frontier models, when given insufficient context, hallucinate rather than abstain at high rates and become more confident as additional but insufficient context is added \citep{joren2025sufficientcontext}. The training interventions that improve a model's measured legal reasoning are, by this evidence, the same interventions that worsen the model's behavior under the conditions of input insufficiency that characterize pro se use. Consequently, current developments suggest the possibility that pro se users will be subject to increasingly higher rates of legal hallucination over time.

If that is the case, it is plausible that the upper bound rises while the lower bound moves more slowly or even in the wrong direction. Current benchmarks measure the upper bound and report it as legal AI progress while the lower bound is unmeasured, rendering the gap between the two invisible. Additionally, it is possible that model robustness to input quality can vary significantly across different frontier models, and the best model for raw legal reasoning may not be the best model for a pro se litigant.

\subsection{The Consequences for Pro Se Litigants}

In contrast to a lawyer reviewing AI-generated outputs and calibrating their reliance accordingly, pro se litigants have a reduced capacity to do the same. On top of the calibration problem, the perceived authority of AI may also increase the credibility the litigant assigns to incorrect outputs, particularly when that output aligns with a folk-legal theory the litigant brought to the conversation or renders realistic-looking hallucinated cases. The result is a compounding harm: the population producing the inputs most likely to elicit model error is also the population least equipped to detect it when an error has occurred.

\subsection{Is There an Embedded Lawyer-Centric Assumption?}

Current benchmark methodology seems to rest on an implicit assumption about how legal AI will be used. Namely, there will be a lawyer or other qualified intermediary supplying the inputs and reviewing outputs. This choice is not incorrect per se. The problem is that while legal AI development is increasingly justified by access to justice goals, the inherited benchmark methodology continues to encode a lawyer-centric vision of deployment. Moreover, the traditional legal system has already long been subject to criticism that it was not built with unrepresented litigants in mind \citep{budzinksi2022prose}. Continued optimization against benchmarks built on this assumption bakes the lawyer-centric configuration into future legal AI development, in a path-dependent way that may be hard to reverse once a generation of models has been trained and selected against these targets.

\section{Evaluation Primer}

In this section, we seek to supplement the theoretical justifications of sections 2 and 3 with a set of empirical observations that suggest why studying the lower bound of legal reasoning in legal benchmarks is a nontrivial pursuit. Using a small sample of multiple-choice questions from the LEXam legal benchmark, we apply distortions along two major dimensions that are common to pro se claimants to observe whether there is degradation in model capability. By doing so, we seek to mimic the manner in which a pro se claimant may interact with an LLM. While this is far from definitive evidence of how model behavior may deviate under pro se conditions and we do not seek to draw substantive conclusions, we hope it serves as a primer to generate discussion in the field as to how legal benchmarks should think about legal reasoning capabilities when the model is engaged in suboptimal manners. By introducing these small experiments, we try to demonstrate by example how we think the legal benchmark literature can incorporate a lower-bound robustness analysis. 

When implementing both distortions, we draw upon multiple choice questions in the LEXam benchmark. For consistency and ease of assessing accuracy, we use a randomized set of 100 English multiple-choice questions from LEXam that have four answer selections each. Due to the limitations we face with testing across multiple families of general purpose models available commercially, we are comparing three different models (GPT-4.1-mini, GPT-4.1-nano, and GPT-4o-mini) within the OpenAI ecosystem. Choosing these three models allows us to compare models within the same size class and also between different size classes. We then compare model accuracy on these multiple choice questions with a distorted version of the questions. Model accuracy is defined as the raw number of questions that the model answers correctly over the total number of sampled questions. While we understand model abstention in producing a NaN or empty answer is preferable to a confident but incorrect answer, we have focused solely on affirmatively correct answers as we are limited by the small sample size of this primer and recognize this would be an interesting area of future study in designing robust legal benchmarks in the future.

To maintain consistency with the original LEXam benchmark, the original LEXam prompts that outline the parameters of task (such as encouraging logical thinking, specifying output format, etc.) are untouched.

\subsection{Typo Perturbation}

For the first distortion dimension, we start with the simple case of typos. We consider typos of three kinds: one-character deletions ("legal" may become "lgal"), one-character swaps ("legal" may become "elgal"), and keyboard-adjacent swaps ("legal" may become "lehal"). We insert these typos with equal probability at varying frequencies. Consider the following clean and distorted version of a sample multiple-choice question, where one typo is inserted every four words:

\inlinepaperfigure{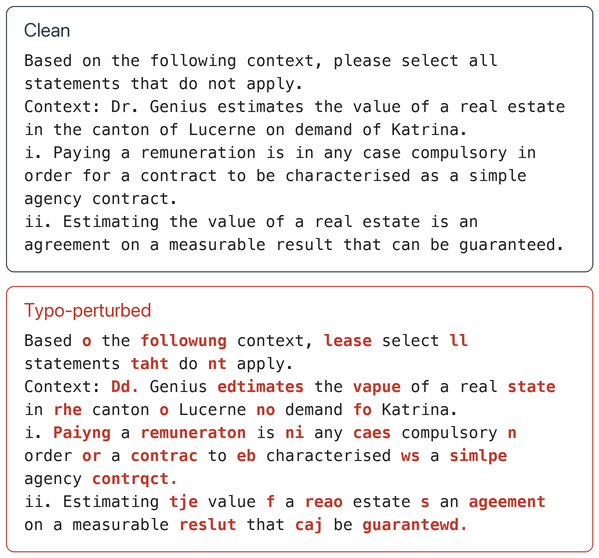}{1}{Clean and Typo-Perturbed LEXam Multiple-Choice Question}{fig:typo-example}

We test typo distortions at three different frequencies to mimic increasing orders of severity (every 2 words, every 3 words and every 4 words) across three different models. The results are as follows:

\inlinepaperfigure{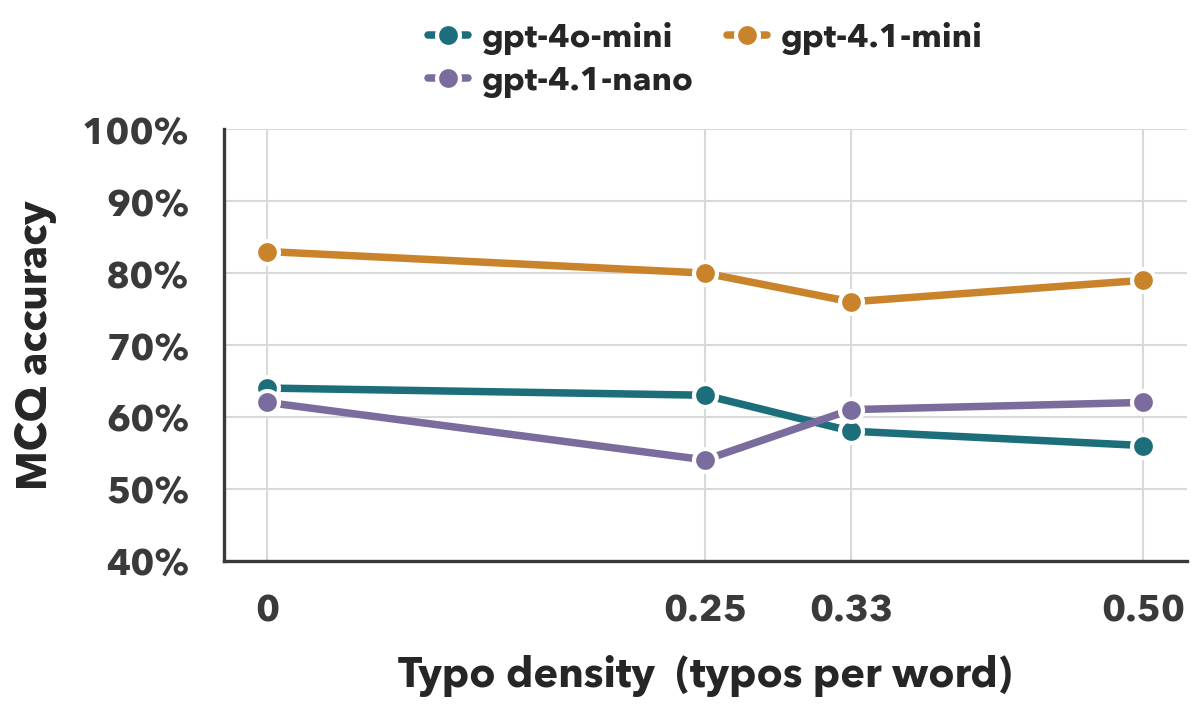}{1}{Model Accuracy Across Typo Perturbation Frequencies}{fig:typo-density}

For the small sample of questions we tested, we observe a general degradation in the legal reasoning ability of all models. Moreover, Figure 2 shows that even simple modifications such as single-character typos can generate an uneven impact across models. The preliminary evidence suggests that the effect of typos may even be large enough to introduce performance rank instability. This would be in line with general frontier models benchmarks, where Alzahrani et al. found that minor perturbations to multiple-choice benchmarks can move models up or down by as many as eight positions in leaderboard rankings \citep{alzahrani2024benchmarks}. Additionally, as typo density increases, not all models in the sample exhibit strictly monotonic degradation, suggesting potential model-by-model discrepancies in how they allocate token resources as it observes further degrading input quality. Models do, in fact, observe typos explicitly. Even when the prompt asks specifically for a multiple choice answer and nothing else, gpt-4.1-nano specifically began several of its answers with "Likely a typo..." before proceeding with the answer. Larger sample sizes and more runs across a wider array of models for teams with more substantial compute resources may help draw firmer conclusions on the strength, significance and stability of these changes. 

\subsection{Context Dilution Perturbation}
The second distortion dimension we test relates to the LLM's difficulty in finding relevant facts in long context windows as discussed in section 3.1. We perform two flavors of this perturbation. In the first setting, we embed the original multiple choice question between two blocks of irrelevant filler sentences. To avoid substantively changing any particular fact of the question, we draw on a bank of sentences that are written in the third-person and describe beautiful scenery. We chose substantively irrelevant filler deliberately as a best-case scenario. A scenic description is distant from any legal question and contains no content that could plausibly be mistaken for operative facts. Any degradation we observe under this benign condition would therefore be a conservative estimate. We predict even higher degradations for partially on-topic padding consistent with Appendix B in Liu et al. The padding sentences include no instructions, no legal terminology and no numbers. We prepend a fixed number of these semantically inert filler sentences before the question and then append a fixed number following the question, producing a "padding sandwich" that preserves the question's content verbatim but dilutes its salience within the prompt. See Figure 3 for a diagram to illustrate the design. 

\inlinepaperfigure{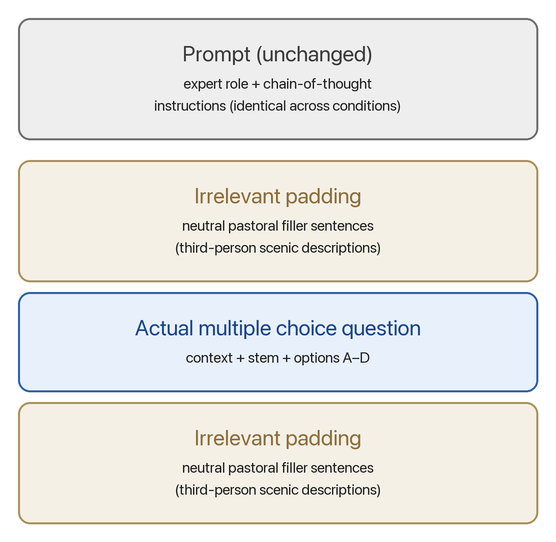}{0.8}{Padding-Based Context Dilution Design}{fig:padding_diagram}

This design is intended to isolate the model's robustness to task-irrelevant distractor context and its ability to locate and attend to the actual question when surrounded by uninformative prose, which seeks to model the pro se inclusion of non-legally relevant facts. To make sure that the model cannot rely on formatting cues to identify relevant content, we flatten the entire structure displayed in Figure 3 by removing all new lines. This forces the frontier models to rely solely on the semantic relevance of the prompt. We evaluate this for three models in two settings of variable padding: 10 sentences per filler block and 30 sentences per filler block. The results are displayed below: 
\inlinepaperfigure{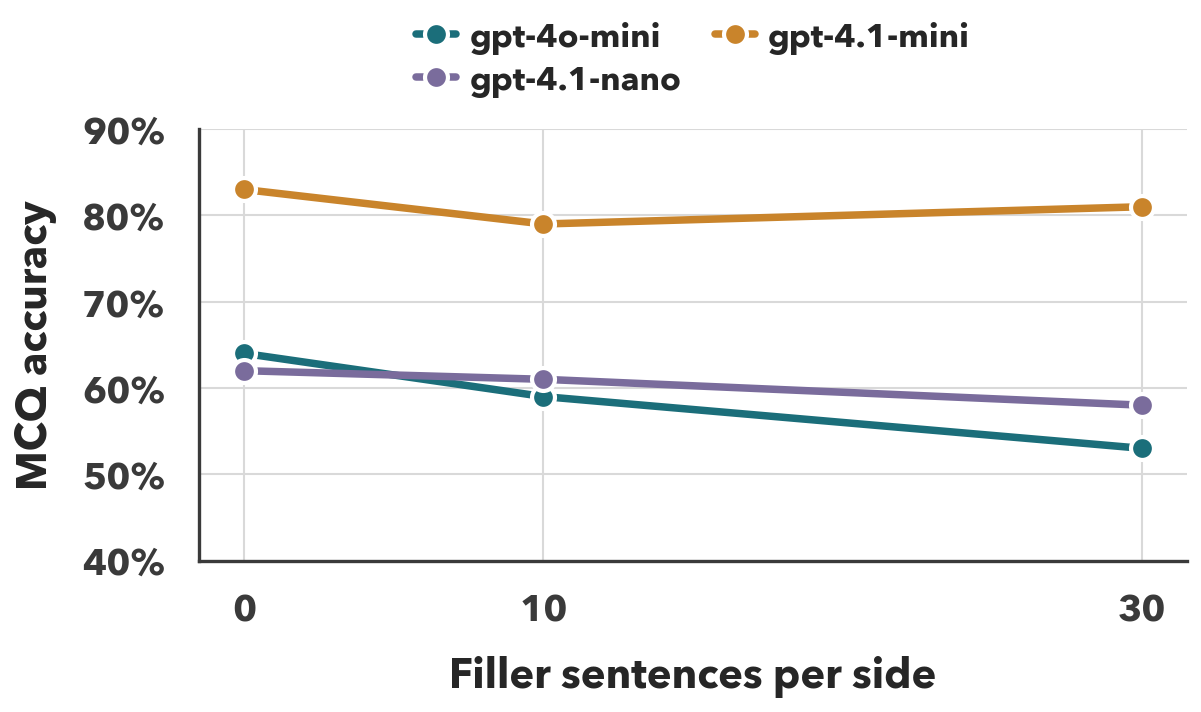}{1}{Model Accuracy Under Padding-Based Context Dilution}{fig:padding_results}

In the second version of this perturbation, we intersperse the content of the question throughout the whole prompt. We do so by inserting two filler sentences between each question sentence as well as between each answer choice. By not including the question as a contiguous chunk of the prompt, this setting tests the ability of the model to repeatedly identify or discard sentences depending on if they serve as actual context for the question. The following diagram gives an illustrative depiction of this. 

\inlinepaperfigure{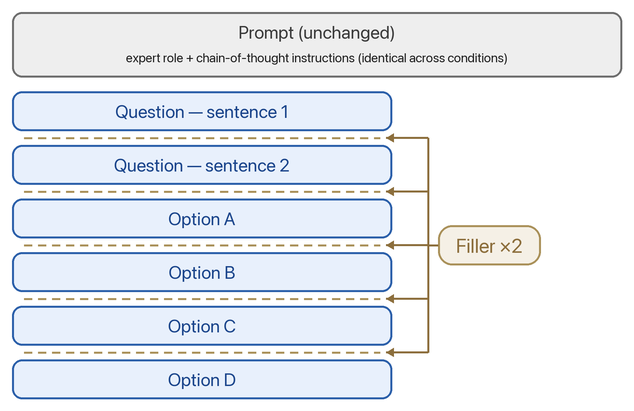}{1}{Interleaved Context Dilution Design}{fig:interleave_diagram}

We used three models to test the effect of interleaving filler sentences. The results are as follows. 
\inlinepaperfigure{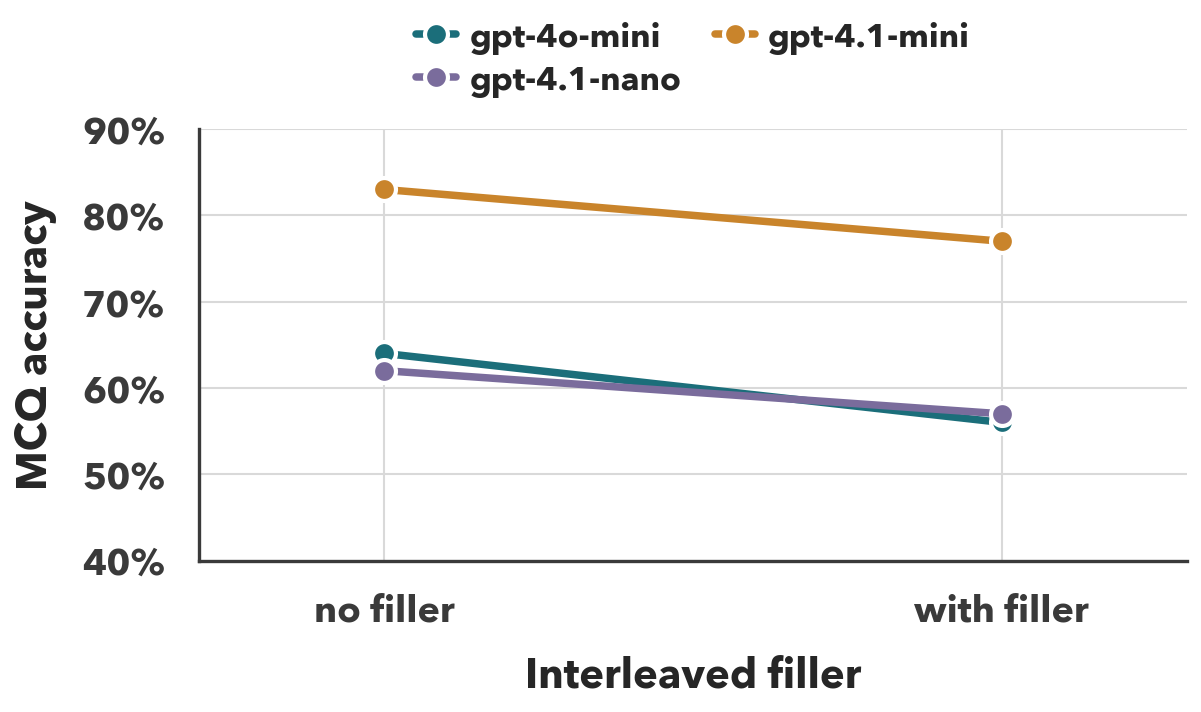}{1}{Model Accuracy Under Interleaved Context Dilution}{fig:interleave_results}

In both flavors of the context dilution perturbation, we observe similar trends to the typo perturbation. Results from the general LLM sensitivity literature seem to translate similarly in affecting legal reasoning of LLMs. By inserting filler sentences of beautiful scenery, we also note that we are purposely estimating a best-case scenario and still observing degradation. This is because in researching long-form context processing, Liu et al. found that inserting completely unrelated context filler yields the best model performance, since models expend less energy differentiating the non-material content \citep{liu2024lost}. In the case of pro se litigants, their prompts may be filled with substantially more relevant, yet nonmaterial facts, which we suspect will cause even more material degradations. 

Through these experiments, we also have reason to suggest that the effect of input degradation may induce rank instability and serve as an area of further research. In all three graphs, we note that the ranking of the models shifted from the clean to distorted version of the data. Moreover, even though all models tested in this small sample were developed by OpenAI, they exhibit different levels of robustness, sometimes in opposite directions when faced with the exact same prompt changes. 

Together, these two types of perturbations help give some empirical backing to the theoretical structure of sections 2 and 3, further justifying the need for future legal benchmarks to specifically investigate how similar perturbations can affect frontier models generally and how that might implicate access to justice. We next discuss general directions the legal benchmark literature should proceed in. 

\section{Areas of Future Research}
\subsection{How Future Benchmarks can Evolve}

There are several empirical questions that we deem critical to answer for access to justice. We frame these as questions rather than as prescriptive suggestions because the answers are not yet known.

The first question is whether the gap exists in the legal domain at the magnitude the general ML literature suggests. While section 3 establishes the conditions under which LLMs are known to degrade and section 5 offers a small experimental demonstration on LEXam, there has not been a study so far that measures performance under controlled pro se-styled input degradation across multiple legal domains. The second question is which models exhibit the smallest gap. We cannot assume that performance under lay input conditions will track performance under expert input conditions across models. A model that scores highest on lawyer-mediated benchmarks may not be the model that performs best under pro se input conditions, and the model that minimizes the gap the most may not be the same one that maximizes either upper or lower bounds. Different models may be robust to different types of perturbations that pro se claimants are prone to. The third question concerns the mechanisms behind whatever gap exists. Although the general ML literature offers candidate explanations such as long context windows or omission and abstention, their relative contributions in the legal domain have not been measured. Knowing the dominant mechanism matters for how the gap might be addressed.

\subsection{What Further Research Enables}

If these questions are addressed, the following outcomes become possible. First, the legal challenges that pro se users will face in an AI-empowered legal system would become measurable as opposed to anecdotal. The current discourse on AI-assisted pro se litigation relies on isolated reports of failure. A measurement infrastructure that captures the lower bound would convert these anecdotes into trend data and allow the field to track how access to justice metrics are evolving as models develop. Second, legal service providers would have a basis for recommending models to the users they serve and have a better basis for educating pro se users on how to best use AI models for legal questions. Third, model developers and legal tech firms would face more of an incentive to build models that perform well under pro se input conditions. The current incentive structure rewards benchmark performance because that is what is visible. If lower-bound performance were equally visible, the optimization target could shift. Fourth, the lawyer-centric assumption embedded in current evaluation practice would become contestable in a way it currently is not.

\subsection{Final Remarks}

Better legal reasoning does not automatically translate into better access to justice. A frontier model that improves on LegalBench, LEXam, or LegalBench-RAG may become a better tool for the lawyers and legal professionals who can supply the inputs those benchmarks assume \citep{guha2023legalbench,fan2025lexam,pipitone2024legalbenchrag}. It does not, by virtue of that improvement, become a better tool for pro se litigants. The mechanism that produces benchmark gains is not the mechanism that determines deployment usefulness for users who cannot supply well-formed inputs. The conflation of the two is currently absorbed by the field as a default rather than examined as an assumption.

Focusing legal benchmarks on lower-bound performance would reward developers for building models that handle realistic deployment inputs and make the access to justice claims that motivate much of legal AI research empirically falsifiable rather than rhetorically convenient. The methodology to do this exists in the broader ML literature. What remains is to bring it into the legal evaluation infrastructure. The experimental section of this paper offers a small demonstration of one way to do that. We do not claim it is the final form of lower-bound legal evaluation. We only state that it is one tractable form and the broader project of measuring how legal AI performs for the pro se population is overdue.

\bibliographystyle{icml2026}
\bibliography{references}

\end{document}